# Magnetic properties and spin dynamics of 3*d*-4*f* molecular complexes

P. Khuntia[1,2*], M. Mariani[2], A. V. Mahajan[1], A. Lascialfari[2,3], F. Borsa[2,4], T. D. Pasatoiu[5], M. Andruh[5]

[1] Department of Physics, Indian Institute of Technology Bombay, Powai, Mumbai-400076, India

[2] Department of Physics "A. Volta" and CNISM, Università di Pavia, I-27100 Pavia, Italy

[3] Department of Molecular Sciences Applied to Biosystems, Università degli Studi di Milano, I-20134 Milan, Italy

[4] Ames Laboratory and Department of Physics and Astronomy, Iowa State University, Ames, Iowa 50011, USA

[5] Inorganic Chemistry Laboratory, Faculty of Chemistry, University of Bucharest, Str. Dumbrava Rosie 23, 020464 Bucharest, Romania

**Abstract**

We present the magnetic properties of three recently synthesized binuclear molecular complexes [NiNd], [NiGd] and [ZnGd] investigated by dc magnetization and proton nuclear magnetic resonance (NMR) measurements. The high-temperature magnetic properties are related to the independent paramagnetic behavior of the two magnetic metal ions within the binuclear entities both in [NiNd] and [NiGd]. On lowering the temperature, the formation of a magnetic dimer, with a low-spin ground state due to antiferromagnetic interaction ($J/k_B \approx -25$ K) between $Ni^{2+}$ and $Nd^{3+}$, is found in the case of [NiNd], while in [NiGd] a ferromagnetic interaction ($J/k_B \approx 3.31$ K) between the magnetic ions leads to a high-spin ($S = 9/2$) ground state. The temperature dependence of the proton nuclear spin lattice relaxation rate $T_1^{-1}$ in [NiNd] is driven by the fluctuation of the hyperfine field at the nuclear site due to relaxation of the magnetization. At high temperature the independent $Ni^{2+}$ and $Nd^{3+}$ spins fluctuate fast while at low temperature we observe a slowing down of the fluctuation of the total magnetization of the dimer because of the insurgence of antiferromagnetic spin correlations. The relaxation mechanism in [NiNd] at low temperature is interpreted by a single, temperature dependent, correlation frequency $\omega_c \propto T^{3.5}$, which reflects the life time broadening of the exchange coupled spins via spin-phonon interaction. The proton NMR signal in [NiGd] could be detected just at room temperature, due to the shortening of relaxation times when T is decreased. The magnetic properties of [ZnGd] are the ones expected from a weakly interacting assembly of isolated moments except for anomalies in the susceptibility and NMR results below 15 K which currently cannot be explained.

## I. INTRODUCTION

Polynuclear complexes containing identical or different metal ions are of increasing interest in molecular magnetism, for the study of the exchange interactions between the paramagnetic metal ions via organic ligands and spin dynamics. In these systems the intermolecular interactions are weak in comparison to intramolecular superexchange interactions because of the presence of non-magnetic organic ligands, so that molecules behave independently one each other. Therefore, their magnetic properties and spin dynamics are an ensemble average of the ones of single molecule. As a result, molecular clusters are excellent zero dimensional model systems for the investigation of nanoscopic and mesoscopic magnetism [1]. With access to different chemical techniques, it is possible to engineer a variety of molecular complexes having different types and number of magnetic atoms in each molecule. Furthermore, due to the small number of magnetic atoms involved, it is often possible to analytically or numerically determine the ground state and excited states properties of such materials. Thus, magnetic clusters offer a fascinating arena in which one can study fundamental physical phenomena using both experiment and theory. Among these, complexes such as $Cr_8$, $Cu_6$, $Cu_8, V_6$, $Fe_{30}$, $Fe_8$, $Mn_{12}$ and many more have been synthesized and investigated in the last couple of decades [1]. A variety of features such as quantum tunneling of magnetization in $Mn_{12}$ and $Fe_8$, discrete magnetic excitation spectrum, transition from quantum to classical physics, and other dynamical effects at very low temperature at the nanoscale in molecular clusters have drawn a lot of attention [1,2, 3]. Recently, considerable attention has been paid to 3d-4f heteronuclear complexes, behaving as molecular magnets, because of their rich physical properties and potential applications [1,6-13]. Magnetic properties of 3d-4f complexes are expected to be different from the transition metal ions because of the large angular momentum of the lanthanide ions. The advantage of utilizing lanthanide ions in the synthesis of single molecule magnet (SMM) is that they offer an energy barrier to the relaxation of the magnetization due to the combination of a large ground-state spin-multiplicity and a strong Ising-type magnetic anisotropy [1,4,5]. Additional features, such as electropositive nature of rare-earth metals with a uniformity of chemical properties and the ability to choose a rare-earth metal of a particular ionic size can lead to the possibility of tuning the properties of these compounds [1,6-13].

NMR is a powerful local probe to investigate static and dynamic properties of molecular clusters. NMR studies on 3$d$-transition metal complexes exhibiting SMM behaviour have been well established, but there has been no NMR investigation on 3$d$-4$f$ molecular complexes so far. We report here on our results of proton NMR measurements coupled with bulk magnetic measurements using a SQUID magnetometer, on three different 3$d$-4$f$ binuclear molecular complexes [NiNd], [NiGd] and [ZnGd]. The paper is organized as follows. In Sec. II we summarize the synthesis of the molecules and their crystal structure. In Sec. III we give a short description of the experimental methods utilized, while the results and their analysis are presented in Secs. IV and V for the magnetization and the NMR experiments respectively. The conclusions are drawn in Sec. VI.

## II. SYNTHESIS AND CRYSTAL STRUCTURE

The details of synthesis and of crystallographic measurements for [NiGd] and [NiNd] complexes have been discussed elsewhere [14]. Within the Ni(II)-Nd(III) system, two slightly different binuclear entities were observed within the same crystal. The synthesis and crystal structure of the [ZnGd] complex have been described elsewhere [15]. Here, we only mention that the three compounds have been obtained following the same general procedure: the mononuclear complexes, [Ni(H$_2$O)$_2$(valpn)] or [Zn(H$_2$O)(valpn)], have been reacted in acetonitrile with Ln(NO$_3$)$_3$·6H$_2$O in a 1 : 1 molar ratio (Ln = Gd, Nd) (H$_2$valpn is the Schiff base resulted from the condensation of $o$-vanillin with 1,3-propanediamine in a 2 : 1 molar ratio). The perspective views of the two compounds are shown in Figures 1 and 2. Their structures consist of binuclear entities, with the Ni(II)/Zn(II) ions connected to the Ln(III) by phenoxo-bridges. The nickel ions within the [NiGd] and [NiNd] complexes are hexacoordinated with an octahedral geometry. The two [NiNd] complexes differ through the coordination environment of the Ni(II) ions: in one complex the apical positions of the Ni(II) ions are occupied by two CH$_3$CN molecules, while in the second one the apical positions are occupied by an aqua ligand and an acetonitrile molecule.

The zinc ion within the [ZnGd] complex shows a square pyramidal geometry. Both Ni(II) and Zn(II) ions are hosted into the N$_2$O$_2$ compartment of the Schiff-base ligand, while the lanthanides occupy the open, large compartment.

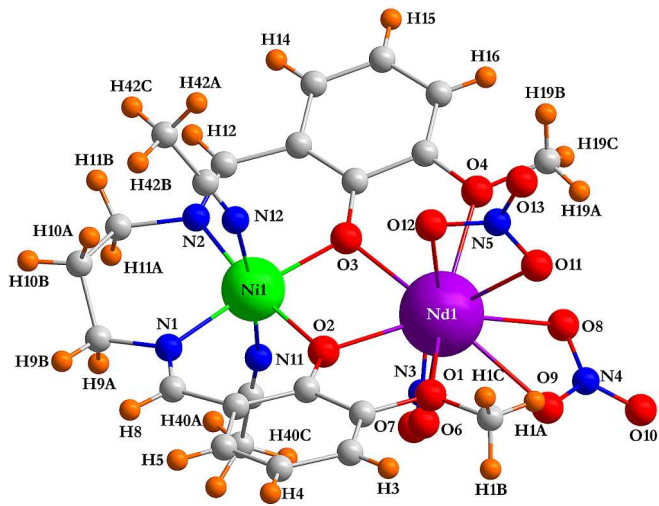

Fig. 1. (Color online) A view of the molecular structure for the [Ni(CH$_3$CN)$_2$(valpn)Nd(O$_2$NO)$_3$], molecular units (in short [NiNd])

[NiNd]: Each unit of the [NiNd] molecule consists of one Ni$^{2+}$ ($S_{Ni}$ = 1) ion and one Nd$^{3+}$ ($S_{Nd}$ = 3/2, $L_{Nd}$ = 6, $J = L$-$S$ = 9/2) ion. The intra-molecular magnetic interaction between Ni and Nd is presumably via O, forming the [NiNd] dimer.

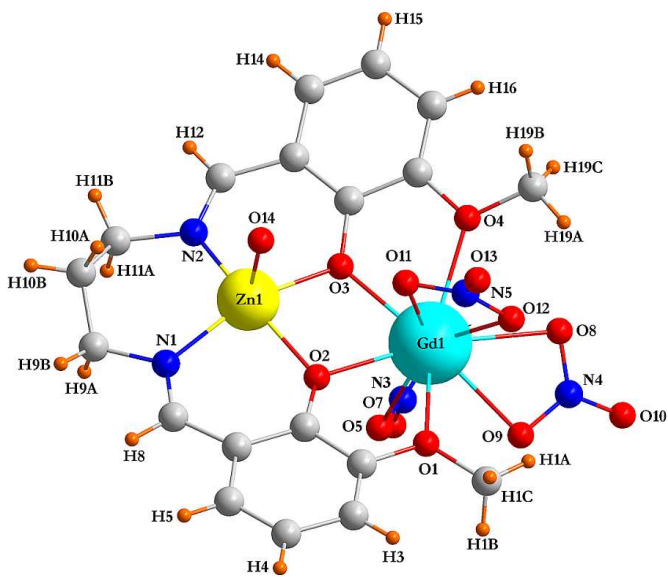

Fig. 2. (Color online) A view of the molecular structure of [Zn(valpn)(H$_2$O)Gd(O$_2$NO)$_3$] (in short [ZnGd]). [ZnGd] is isostructural with [NiGd].

**[ZnGd]:** Gd$^{3+}$ ($S = 7/2$, $L = 0$) is the only magnetic ion in [ZnGd], and each Gd$^{3+}$ is surrounded by many inequivalent protons. Each Gd is far away from another Gd and is separated by organic ligands. Thus, only a weak intermolecular interaction (of dipolar origin) is expected.

**[NiGd]:** Each unit of [NiGd] consists of one Ni$^{2+}$ ($S_{Ni}$ =1) and one Gd$^{3+}$ ($S_{Gd}$ = 7/2) and the interaction between Ni and Gd is possible via the phenoxy groups. So [NiGd] can be regarded as a heterodinuclear Ni$^{2+}$–Gd$^{3+}$ complex.

## III. EXPERIMENTAL DETAILS

The temperature dependence of the magnetic susceptibility, ($\chi = M/H$) in the temperature range $2 \leq T \leq 300$ K in different magnetic fields and the magnetization isotherms ($M$ vs. $H$) for $0 \leq H \leq 50$ kOe at 2 K were measured with a Superconducting Quantum Interference Device (SQUID) magnetometer (Quantum Design, Inc.). The raw data were corrected for the contribution from the sample holder. The diamagnetic contribution from the ion cores ($\chi_{core} \sim 10^{-4}$ cm$^3$/mol) [16] was subtracted from the raw magnetic susceptibility data before analysis.

NMR measurements on polycrystalline samples of [NiNd] and of [ZnGd] including proton absorption spectra, spin-lattice relaxation time ($T_1$), and spin-spin relaxation time ($T_2$) were performed with a standard Tecmag NMR spectrometer using short $\pi/2$ frequency ($rf$) pulses (2.2-3.4 $\mu s$) in the temperature range $5 \leq T \leq 300$ K for an applied magnetic field of $H = 15$ kOe. Fourier transform of the spin echo signal was taken to obtain the NMR absorption spectra keeping the external field constant. A standard solid echo ($\pi/2$-$\pi/2$) pulse sequence was used to perform the spin-spin relaxation ($T_2$) measurements. The proton NSLR ($T_1^{-1}$) measurements were performed by observing the recovery of the nuclear magnetization (solid spin echo) following a comb of saturating $rf$ pulses in order to obtain the best saturation of the nuclear magnetization.

## IV. RESULTS OF MAGNETIZATION MEASUREMENTS

### A. Magnetic susceptibility

The variation in χ with $T$ for [NiNd] is shown in Fig. 3. The inset shows the temperature dependence of the product of magnetic susceptibility times the temperature, a quantity that is proportional to the square of the effective magnetic moment of the molecule.

In the molecular-field approximation(MFA), the Curie constant is given by $C = \frac{N_A g^2 \mu_B^2 S(S+1)}{3k_B}$ for a 3$d$-ion with quenched orbital angular momentum and $C = \frac{N_A g_J^2 \mu_B^2 J(J+1)}{3k_B}$ for a 4$f$-ion where $N_A$ is the Avogadro number, $g_J$ is the Lande' $g$ factor with $g_J = 1 + \frac{J(J+1) - L(L+1) + S(S+1)}{2J(J+1)}$, $\mu_B$ is the Bohr magneton, and $k_B$ is the Boltzmann constant. Thus, one can write $\chi_m T = \frac{\mu_{eff}^2 N_A}{3k_B}$, where $\chi_m$ is the molar susceptibility and $\mu_{eff}$ is the effective magnetic moment of each molecule. For Ni$^{2+}$ ($S = 1$, $g = 2$) the free ion effective moment is 2.82 $\mu_B$, while for Nd$^{3+}$ ($S = 3/2$, $L = 6$, $J = 9/2$, $g_J = 8/11$) the effective moment is 3.62 $\mu_B$. Therefore, if the two ions contribute independently to the susceptibility one expects that the effective moment is 4.59 $\mu_B$ and $\chi_m T = 2.65$ cm$^3$/mol. As can be seen from the inset in Fig. 3 the experimental value of $\chi_m T$ does approach the free-ion value at high temperatures. The decrease in $\chi_m T$ between 300 K and 10 K is due to the depopulation of the Stark levels of the single Nd(III) ion. The drop of $\chi_m T$ to about 0.4 cm$^3$ K mol$^{-1}$ at even lower temperatures is due to an antiferromagnetic Ni(II)-Nd(III) coupling. Using molecular field approximation (a crude approximation), the exchange couplings between 3$d$-4$f$ moments from the magnetization data taken on polycrystalline samples can be estimated. The antiferromagnetic exchange interaction between Ni and Nd was found to be $J_{NiNd}/k_B \approx -25$ K for NiNd.

This is supported by Goodenough-Kanamori rules, [17,18] which suggest an antiferromagnetic interaction between Ni$^{2+}$ and Nd$^{3+}$.

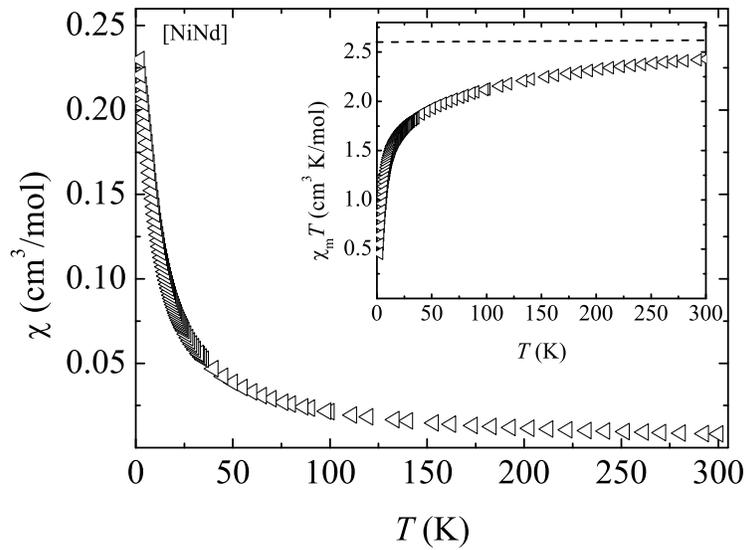

Fig. 3. Temperature dependence of magnetic susceptibility ($\chi$) in [NiNd] in an applied magnetic field of 15 kOe. The inset shows $\chi_m T$ vs. $T$ which corresponds to plotting the square of the effective moment vs. $T$. The dashed line is the independent free ion contribution (see text).

In contrast, for [NiGd], the $\chi T$ vs $T$ data (inset in Fig. 4) shows a positive deviation from the free-ion value (8.93 cm$^3$ K/mol) at high temperatures.

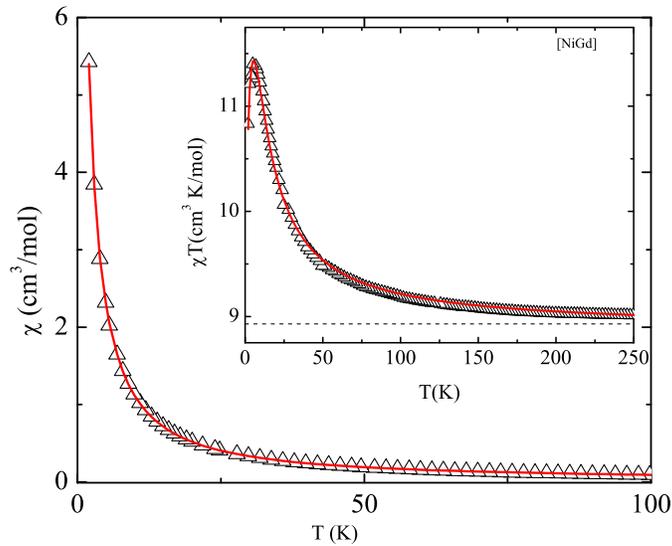

Fig. 4. (Color online) Temperature dependence of $\chi$ vs. $T$ in $H = 1$ kOe. The inset shows the temperature dependence of $\chi_m T$ and the solid line is a fit as discussed in the text. The dashed line represents the independent free ion contribution (see text).

The experimental magnetic susceptibility data in [NiGd] can be interpreted by an effective Hamiltonian, $H = -J_{NiGd} S_{Ni} S_{Gd} + D_{Ni}[S_{zNi}^2 - S_{Ni}(S_{Ni}+1)/3]$, where $D_{Ni}$ is the zero field splitting parameter. The best fit (see Fig.4) was obtained with $J_{NiGd}/k_B$=3.31 K, $D_{Ni}/k_B$= 10.79 K and g=2. So, the exchange interaction between Ni and Gd is ferromagnetic via phenoxo bridges. It may be noted the intermolecular interaction is very weak and antiferromagnetic i.e., $J'_{NiGd}/k_B$ = -0.08 K. These exchange couplings also supported by Goodenough-Kanamori rules [17,18].

For [ZnGd], on the other hand, $\chi(T)$ follows a Curie law over, nearly, the full temperature range with $C = 8.52$ cm$^3$ K/mol (see Fig. 5) which is comparable to the value for free Gd ions (7.87 cm$^3$ K/mol).

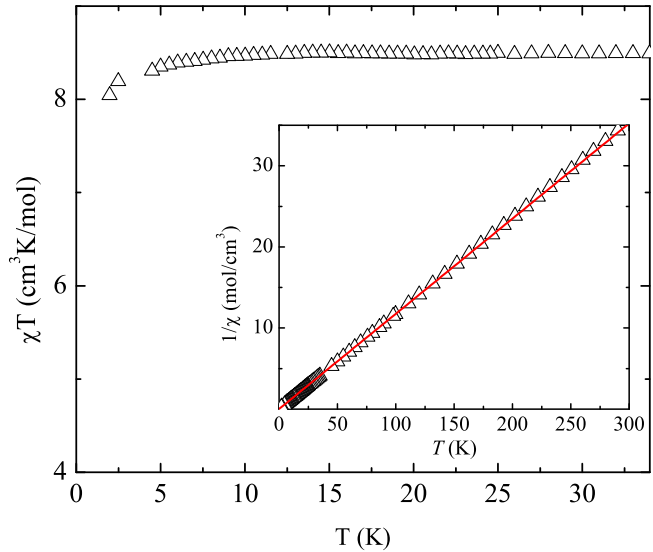

Fig. 5. (Color online) Temperature dependence of $\chi_m T$ in the low temperature region in [ZnGd] in $H = 3.5$ kOe. (b) The inset shows the temperature dependence of $\chi^{-1}$ in 3.5 kOe, and the solid line is the Curie fit.

The slight decrease of $\chi_m T$ below 10 K indicates, possibly, the development of inter-cluster antiferromagnetic interactions among the Gd ions.

Here, it is worthwhile to mention that the electronic level of the ground state multiplet with angular momentum $J$ in the 4$f$ ions is split by strong crystal field anisotropy, which is in contrast with that of the 3$d$ ions. Additionally, the magnetization data also tuned by the local contributions of the ligand field on to the 4$f$ moments and spin-orbit couplings. So it is rather difficult to convincingly estimate the exchange constants and anisotropic parameters from the magnetization data taken on polycrystalline samples. The low temperature magnetization measurements on high quality single crystals are required for quantitative estimation of exchange couplings and anisotropic parameters. However, the rough estimation of exchange couplings from bulk magnetization data captures the main physical features of the molecules presented here.

### B. Magnetization isotherms at 2K.

In order to get information about the ground state properties of these molecular complexes, we have performed magnetization measurements vs external magnetic field at fixed temperatures. The simplest case is the [ZnGd] system since each molecule contains a single magnetic moment, the results of which are shown in Fig.6.

As can be seen, [ZnGd] behaves like a paramagnet and more specifically, one can categorize it as a single moment molecule paramagnet.

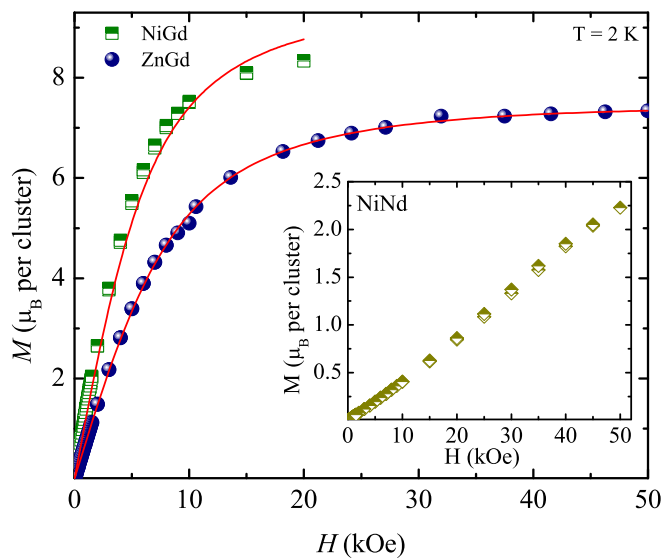

Fig.6. (Color online) Magnetization isotherms at 2 K in [NiGd] and [ZnGd]. The inset shows the magnetization isotherm in [NiNd] at 2K. The solid lines are the expected behavior of the magnetizations following Brillouin functions at $T = 2$ K in [NiGd] ($S =9/2$ and $g = 2$) and [ZnGd] ($S =7/2$ and $g = 2$).

The field dependence of the magnetization of [NiGd] at 2 K can be explained assuming paramagnetic behavior of $S = 9/2$ entities. So, at low temperature, the molecular cluster [NiGd] exhibits a high spin ground state with $S = 9/2$ originating from the ferromagnetic coupling between $Ni^{2+}$ and $Gd^{3+}$.

The magnetization data in NiNd indicate the existence of an antiferromagnetic intradimer interaction, which implies a dimer ground state with an effective magnetic moment smaller than the two independent moments. However, the implications of this result in terms of the structure of the dimer ground state are not presently understood.

## V. NMR RESULTS

### A. $^1$H NMR Spectra

Proton NMR spectra for [NiNd] were taken as a function of frequency at constant applied magnetic field at various temperatures. $^1$H NMR spectra broaden progressively with lowering the temperature due to the presence of many inequivalent proton sites contributing towards the overall broadening of the line shape. The results are shown in Fig.7.

The shape and width of the $^1$H NMR line is governed by the nuclear-nuclear dipolar interaction, and the hyperfine interaction of the protons with $Ni^{2+}$ and $Nd^{3+}$. The first interaction generates a temperature and field independent broadening [19, 20]. The second contribution to the line width is associated with the average static component of the magnetic moment.

Hence, in the simple Gaussian approximation, the NMR line width is related to the sum of the second moments due to the two different mechanisms as follows [19]:

$$FWHM \propto \sqrt{\langle \Delta \nu^2 \rangle_d + \langle \Delta \nu^2 \rangle_m} \qquad (1a)$$

where $\langle\Delta\nu^2\rangle_d$ is the intrinsic second moment due to nuclear dipolar interactions and $\langle\Delta\nu^2\rangle_m$ is the second moment of the local frequency-shift distribution due to neighboring electronic moments at the different proton sites in the molecule [19,21]. The second moment $\langle\Delta\nu^2\rangle_m$ is proportional to the bulk magnetic susceptibility. The relation between $\langle\Delta\nu^2\rangle_m$ and local $Ni^{2+}$ and $Nd^{3+}$ electronic moments for a dipolar interaction is given by [19,21]:

$$\langle\Delta\nu^2\rangle_m = \frac{1}{N}\sum_R\left(\sum_{i\in R}\langle\nu_{R,i}-\nu_0\rangle_{\Delta t}\right)^2 = \frac{\gamma^2}{N}\sum_R\left[\sum_{i\in R}\sum_{j\in R}\frac{A(\vartheta_{i,j})}{r_{i,j}^3}\langle m_{z,j}\rangle_{\Delta t}\right]^2 \qquad (1b)$$

where $R$ denotes different molecules, $i$ and $j$ refers to different protons and $Ni^{2+}$ and $Nd^{3+}$ ions within each molecule respectively, and $N$ is the total number of protons probed. Here, $\nu_{R,i}$ is the NMR resonance frequency of nucleus $i$ and $\nu_L = (\gamma/2\pi)\,H$ is the bare Larmor frequency. The shift for nucleus $i$ due to the local field generated by the nearby moments j, originated from the difference between aforementioned two resonance frequencies. $A\,(\vartheta_{i,j})$ refers to the angle dependent dipolar coupling constant between nucleus $i$ and moment $j$, and $r_{i,j}$ is the corresponding distance. $<m_{z,j}>$ represents the component of the $Ni^{2+}/Nd^{3+}$ magnetic moment $j$ in the direction of the applied field, averaged over the NMR data acquisition time scale. In the case of a simple paramagnet one expects $\langle m_{z,j}\rangle = \frac{M}{N_A}$, where $M$ is the magnetization and $N_A$ is Avogadro's number.

Thus, if $H$ is the external magnetic field, one can write approximately,

$$\frac{\sqrt{\langle\Delta\nu^2\rangle_m}}{H} = A_z\chi \qquad (2)$$

where $A_z$ is the dipolar coupling constant between protons and magnetic moments of the ions averaged over all protons and all orientations, and $\chi\,(=M/H)$ is the SQUID susceptibility.

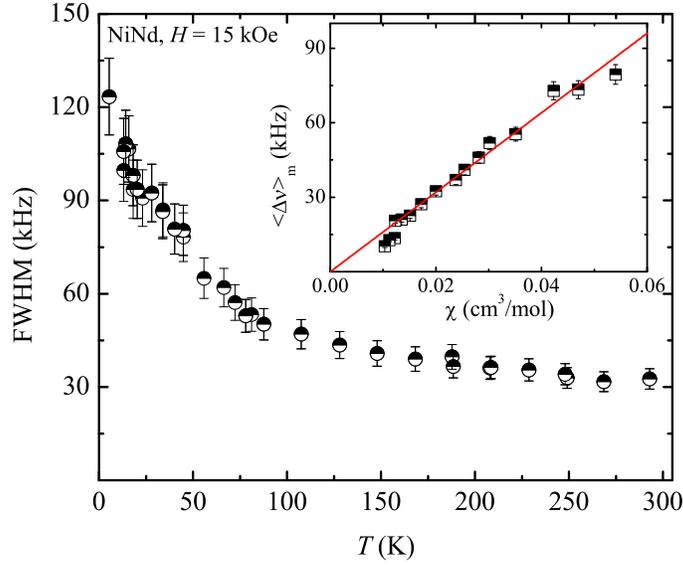

Fig. 7. (Color online) Temperature dependence of the proton full width at half maximum (FWHM ) at $H$ = 15 kOe. The inset shows the magnetic inhomogeneous broadening $\left(\langle\Delta\nu\rangle_m = \sqrt{\langle\Delta\nu^2\rangle - \langle\Delta\nu^2\rangle_d}\right)$ vs $\chi$ in [NiNd] with the fit to Eq. 2 as discussed in the text.

The magnetic contributions to the line width ($\langle\Delta\nu\rangle_m$) are plotted as a function of the bulk magnetic susceptibility (inset of Fig. 7) for [NiNd] with the temperature as an implicit parameter. The linear behavior of $\langle\Delta\nu\rangle_m$ with $\chi$ is interpreted by Eq. 2, and the value obtained from the fit for the average dipolar coupling is $A_z = 1.5\times10^{22}$ cm$^{-3}$. This value of $A_z$ is consistent with the dipolar interaction of protons with the Ni$^{2+}$ and Nd$^{3+}$ magnetic ions at a mean distance of 4 Å. One can conclude that the hyperfine interaction of protons with the Ni$^{2+}$ and Nd$^{3+}$ magnetic moments is mostly of dipolar origin, a scenario which is consistent with a negligible mixing of the $s$-wave functions at the hydrogen site with the $3d$ and $4f$ wave functions, which would lead to a much stronger superexchange coupling [22].

The line width of the $^1$H NMR spectra in [ZnGd] taken in an applied magnetic field $H = 15$ kOe gradually increases from about 45 kHz at 150 K to about 90 kHz at 5 K (not shown here) and is proportional to the magnetic susceptibility only from room temperature down to about 10 K. Below 15 K, there is a deviation of the line width from the linear behavior predicted by Eq. 2, and in the temperature rage 2.5 ≤ T ≤ 1.5 K, one observes a rapid increase in the line width by more than one order of magnitude.

## C. Spin-spin relaxation time ($T_2$), proton NMR signal intensity and wipeout effects

### (a) [NiNd]

First, we will analyze the results for [NiNd]. For spin-spin relaxation time ($T_2$) measurements, the echo integral (which arises from the transverse magnetization) was taken as a function of time delays between two *rf* pulses. The recovery of the transverse magnetization was found to be exponential at all temperatures, so the spin-spin relaxation time was obtained from fits to $M(t) = M_0 \exp(-2t/T_2)$ (see Fig. 8 inset) where $t$ is the delay between two *rf* pulses and $M_0$ is the initial magnetization. The $T_2$ data obtained are shown in Fig. 8.

The transverse relaxation rate $1/T_2$ contains both static and dynamic contributions. In [NiNd], the static part (~0.025 $\mu s^{-1}$) is expected to be constant over a wide range of temperatures, which corresponds to the nuclear dipole-dipole interactions among protons and thus, is proportional to the square root of Van Vleck second moment [23]. The enhancement of $T_2^{-1}$ in an intermediate temperature regime is attributed to the dynamic contribution, which stems from the hyperfine interaction between nucleus and the fluctuating magnetic moments in [NiNd] in their collective ground state. The dynamic spin-spin relaxation rate in the weak collision limit and in the fast motion approximation can be expressed as the spectral density of fluctuating hyperfine field at zero frequency [24],

$$\left(\frac{1}{T_2}\right)_{dynamic} = \gamma_n^2 \langle \delta H_z^2 \rangle \tau(T) = \frac{\gamma_n^2 \langle \delta \mu_e^2 \rangle}{r^6} \tau(T) \qquad (3)$$

where $\delta H_z$ is the local longitudinal fluctuating field originating from a magnetic moment located at a distance $r$ and $\tau$ is the correlation time, which is determined by the dynamics of exchange coupled electronic spins.

Accordingly, the experimental $T_2^{-1}$ data in NiNd are fitted to $1/T_2 = (1/T_2)_{static} + (1/T_2)_{dynamic}$, which yields $\delta H_z = 250 \pm 30$ Oe while the correlation time was assumed to have the temperature dependence $\tau(T) = D\,T^{-3.5}$ with $D = 10^{-4}$ rad/sec. Temperature dependence of the correlation time was assumed in accordance to what found in many molecular clusters [20] and is justified further below.

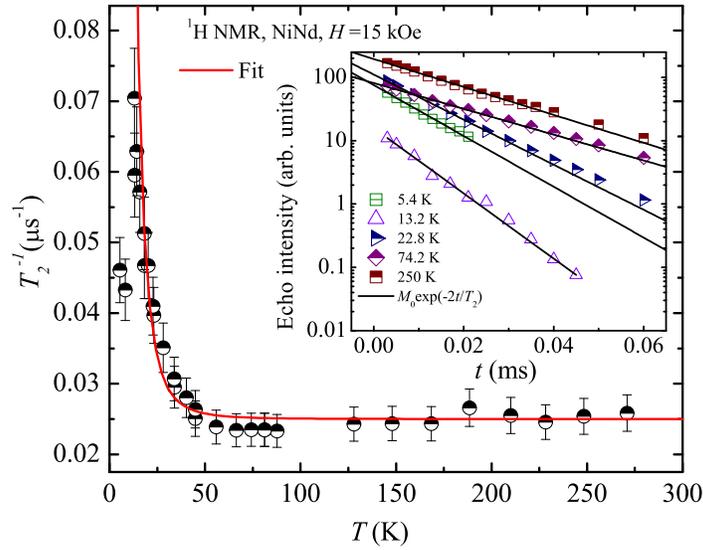

Fig.8. (Color online) Temperature dependence of spin-spin relaxation rate at $H$ = 15 kOe in [NiNd] with a fit according to Eq. 3. The inset shows the recovery of transverse magnetization in [NiNd] at various temperatures. The solid lines in the inset are fits to the exponential behavior of the transverse magnetization as discussed in the text.

The NMR signal intensity at a temperature $T$ can be determined by extrapolating the echo amplitude back to $t = 0$ (in the recovery of the transverse magnetization) and then multiplied by $T$ to compensate for the Boltzmann factor. The normalized echo intensity for protons in [NiNd] at $H$ = 15 kOe is shown below (Fig. 9). We observed a loss in the NMR signal intensity in the intermediate temperature regime which becomes severe at low temperatures, and this is referred to as the wipe-out effect. Here, it is worth mentioning that the loss of NMR signal intensity is a phenomenon also observed in some cuprates [25] and spin-glass materials [26] and, in general, it provides information concerning the hyperfine coupling and spin dynamics of fluctuating magnetic moments at low temperatures.

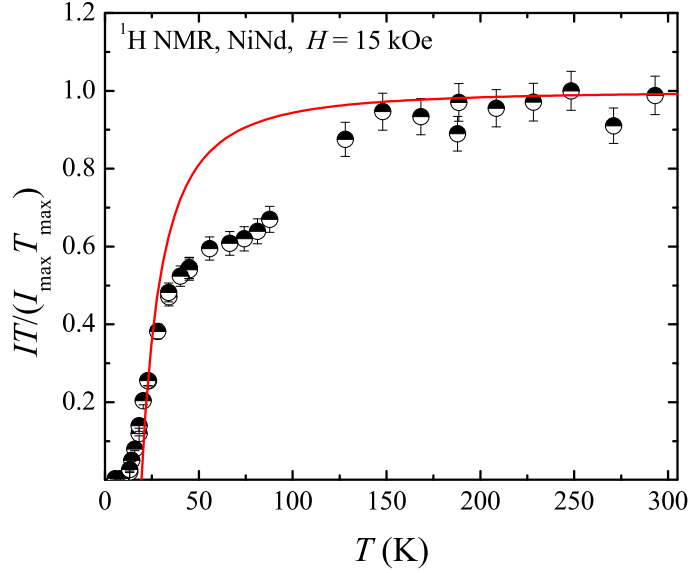

Fig. 9. (Color Online) Temperature dependence of normalized NMR signal intensity $I$ multiplied by temperature. The solid line is a fit with the theoretical model (Eq. 4) discussed in the text.

The origin of the wipe-out effect is attributed to the fact that a fraction of the protons closer to the magnetic $Ni^{2+}$ and $Nd^{3+}$ ions can attain spin-spin relaxation times ($T_2$) shorter than the dead time ($T_2 \ll \tau_d$) of the spectrometer, and the signal from these protons cannot be detected [20, 27]. Only the protons which are outside the critical radius of the magnetic ions contribute towards the signal intensity. The loss of NMR signal intensity is associated with the coupling of nuclei with $Ni^{2+}$ and $Nd^{3+}$ spins and the slowing down of the fluctuations of magnetic moments with lowering the temperature.

A simple theoretical model that explains the loss of proton NMR signal intensity (wipe-out effect) has been applied successfully in the case of many molecular clusters [20, 27, 28]. According to this model the number of protons $n(T)$ that contribute to the NMR signal intensity at each temperature is as follows:

$$\frac{n(T)}{n_0} = 1 - \frac{\gamma_n \sqrt{\langle \delta\mu_e^2 \rangle} \sqrt{\tau_d}}{R^3} \sqrt{\tau(T)} \qquad (4)$$

where $n_0$ = total number of protons in each molecule, $R$ = distance between magnetic ions and proton, $\gamma_n$ is proton gyromagnetic ratio and $\tau_d$ (≈15 μs) is the dead time of the spectrometer. The fit of the normalized NMR signal intensity to Eq. 4 is shown in Fig. 9. We assume $\tau(T) = DT^{-3.5}$ with $D = 10^{-4}$ rad/sec as derived from the fit of $T_2$ in Fig.8; thus, the only fitting parameter in Eq.4 is $\sqrt{<\delta\mu_e^2>}/R^3$ (the local fluctuating hyperfine field experienced at the proton site). The best fit is obtained for $\sqrt{<\delta\mu_e^2>}/R^3 = 390 \pm 50$ Oe. This value can be compared with $\delta H_z = 250 \pm 30$ Oe obtained above from the fit of the $1/T_2$ data. The difference between the two values can be ascribed to the simplicity of the model chosen in interpreting the signal loss data. However, these values are of the correct order of magnitude for the dipolar interaction of protons with the magnetic moments of the $Ni^{2+}$ and $Nd^{3+}$ ions and are similar to values obtained in ferromagnetic and antiferromagnetic molecular clusters [27, 28].

(b) [ZnGd]

Now, we consider the spin-spin relaxation rate and normalized NMR signal intensity in [ZnGd]. The results are shown in Fig 10 .

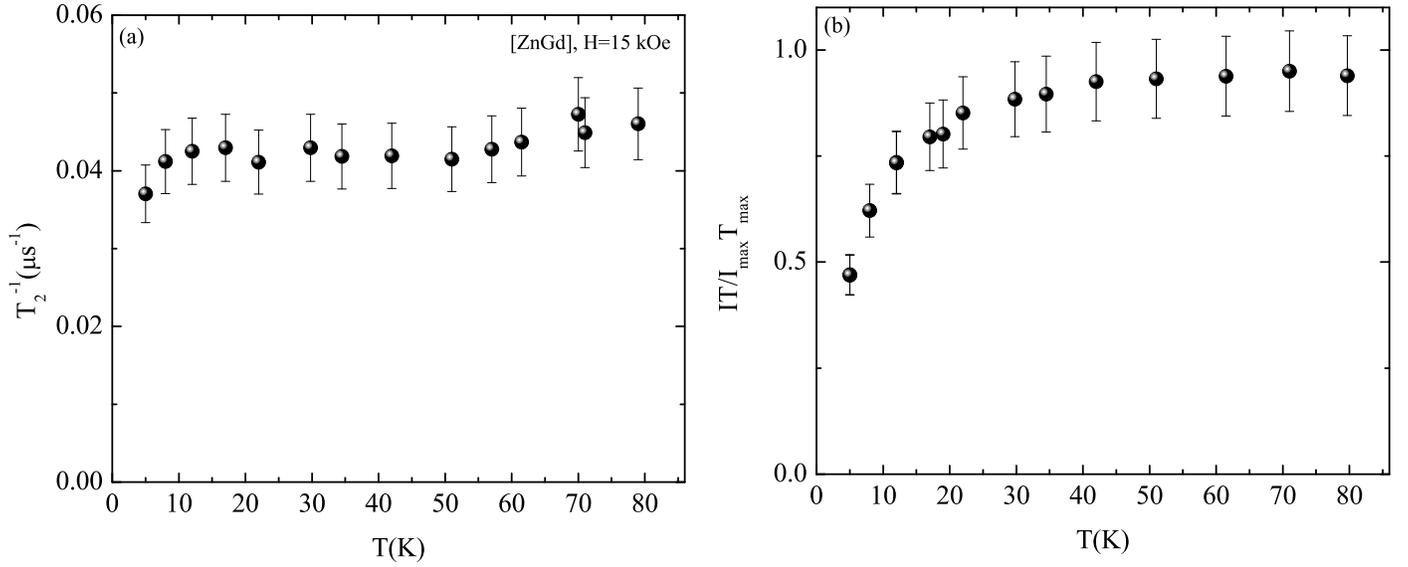

Fig.10. (a) Spin-spin relaxation rate and (b) normalized signal intensity plotted vs. temperature for [ZnGd] in an external field of 15 kOe.

Contrary to the case of [NiNd] both $1/T_2$ and the normalized signal intensity are almost T independent down to about 15 K, indicating no slowing down of the fluctuations as expected for weakly coupled paramagnetic ions (i.e. $Gd^{3+}$ here). The onset of a decrease of NMR signal intensity observed below 15 K is related to the anomaly displayed by the magnetic susceptibility ( Fig.5 ) and to the increase in the NMR line width mentioned earlier.

## C. Nuclear spin- lattice relaxation rate (NSLR): $T_1^{-1}$

### (a) [NiNd]

In order to probe the spin dynamics in [NiNd], the $^1H$ NSLR measurements have been performed on polycrystalline sample. The recovery of the nuclear magnetization was non exponential due to the presence of many inequivalent protons in the irradiated NMR line which is a common feature in molecular clusters [20].

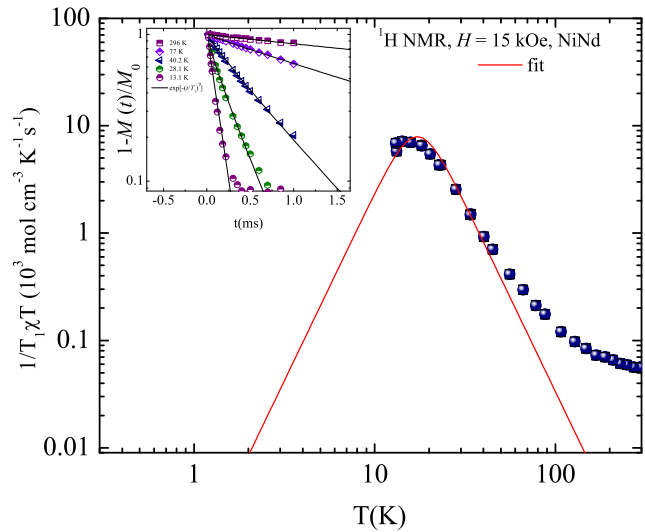

Fig.11. (Color online) Temperature dependence of nuclear spin lattice relaxation rate divided by $\chi T$ at $H$ =15 kOe in [NiNd] with a fit (solid line) as per Eq. 5 discussed in the text. The inset shows the recoveries of longitudinal nuclear magnetization after time delay $t$ following a sequence of saturating pulses. The solid lines in the inset are fits to stretched exponentials.

The non-exponential behavior of nuclear magnetization recoveries at various temperatures can be attributed to the diffusion limited relaxation mechanism [21, 29]. We found that a stretched exponential function $\exp[-(t/T_1)^\beta]$ can fit the recovery of the nuclear magnetization at all temperatures with an almost T independent value of the exponent $\beta$ which is close to 1. (corresponding to a narrow distribution of relaxation times). One also can determine the average relaxation rate, which is dominated by the fast relaxing protons close to the magnetic ions by measuring the initial recovery or tangent at the origin of the semi-logarithmic plot of the recovery of the nuclear magnetization after initial saturation. Alternatively, one can define $T_1$ as the time at which the recovery of nuclear magnetization is reduced to 1/e of the initial value. In the present case, all the above schemes nearly lead to the same $T_1$ value. We have determined $T_1^{-1}$ value from the stretched exponential fit of the recovery of the nuclear longitudinal magnetization where we assume that the measured parameter is close to a weighted average of the relaxation rates of the different protons in the molecule.

The temperature dependence of the proton NSLR in [NiNd] is shown in Fig. 11. The spin dynamics in various temperature regimes in molecular clusters investigated by NMR can give us information regarding the nature of spin correlation function, spin fluctuations, and spin ground states [30, 31]. In the high temperature limit ($k_B T > J$), the magnetic moments in the [NiNd] clusters are weakly correlated, the system behaves like a simple paramagnet, and the NSLR is constant over a wide temperature range as expected [30]. The enhancement of $T_1^{-1}$ in intermediate temperatures and the maximum at a temperature comparable with the exchange coupling constant of the magnetic moments of [NiNd] (as shown in Fig. 11) indicates correlated spin dynamics [20, 31]. The maximum in $T_1^{-1}$ is reached when the correlation frequency (which depends on temperature) is comparable with the nuclear Larmor frequency. The correlation frequency at low temperature is the one that drives the uniform collective fluctuation of the two coupled moments rather than the one driving the fluctuation of the individual $Ni^{2+}$ and $Nd^{3+}$ moments at higher temperatures. Thus, in order to fit the temperature dependence $T_1^{-1}$ in the T region where the maximum is located, we use a simplified version of Moriya theory for NSLR wherein the local hyperfine field at the nuclear site is modulated by the correlation frequency. The correlation frequency accounts for the relaxation mechanism and fluctuation of magnetization and captures the main features of the spin dynamics of most of the magnetic molecules [20, 31, 32, 33]:

$$\frac{1}{T_1} = A\chi T \frac{\omega_c(T)}{\omega_c(T)^2 + \omega_L^2} \qquad (5)$$

where $A$ is the average square of the transverse hyperfine field, $\chi$ is the uniform magnetic susceptibility, $\omega_L$ is nuclear Larmor frequency, and $\omega_c(T) \approx \frac{1}{\tau(T)}$ is the correlation frequency of fluctuating coupled magnetic moments. In analogy with the other molecular nanomagnets we assume for the temperature dependence of the correlation time the power law behavior i.e. $\tau(T) = DT^{-3.5}$, where $D$ is a parameter which is determined by the spin-phonon coupling, is believed to be the mechanism of relaxation of the magnetization of the dimer [31]. The temperature dependence of the proton NSLR ($T_1^{-1}$) in [NiNd] is driven by the magnetic dipole hyperfine interaction between nuclei and fluctuating paramagnetic ions.

By fitting our $1/T_1$ data (Fig. 11) to Eq. 5, we obtain the best fit for $A \sim 10^{13}$ rad$^2$/s$^2$ assuming $D \sim 10^{-4}$ rad/sec, the same obtained above from the temperature dependence of $1/T_2$ (Fig.8) and from the wipe out effect (Fig.9). Thus in the low temperature limit ($k_B T < J$), the spin dynamics in [NiNd] is dominated by the fluctuations of the magnetization of the molecular dimer which are due to spin-phonon interactions [20]. This type of situation is common in molecular clusters but is first of its kind observed in 3$d$-4$f$ complexes.

(b) [ZnGd]

As before, the recovery of the nuclear magnetization in this case was found to be non-exponential and was analyzed as for [NiNd]. The data obtained for the NSLR are shown in Fig 12. The magnitude of $1/T_1$ at high temperatures is about one order of magnitude smaller than in the case of [NiNd]. The high temperature NSLR is almost constant indicating that the relaxation process is driven by nuclear electron dipole-dipole interactions modulated by the paramagnetic fluctuations of the Gd$^{3+}$ ionic moment. For a simple paramagnet at high temperatures one expects $1/T_1 \propto \tau_e$, where $\tau_e$ is the correlation time for the fluctuations of the independent magnetic moment, i.e., the reciprocal of the well known Moriya exchange frequency [33] (different from the correlation time $\tau(T) = D\ T^{-3.5}$ discussed above for the fluctuations of the dimer magnetization of [NiNd] at low temperatures). Comparing the high temperature values of $1/T_1$ in [ZnGd] with the one in [NiNd], one can deduce that the correlation time $\tau_e$ in the former molecule is shorter. This indicates that the weakly interacting Gd$^{3+}$ ions fluctuate faster than the Ni$^{2+}$/Nd$^{3+}$ coupled ionic moments.

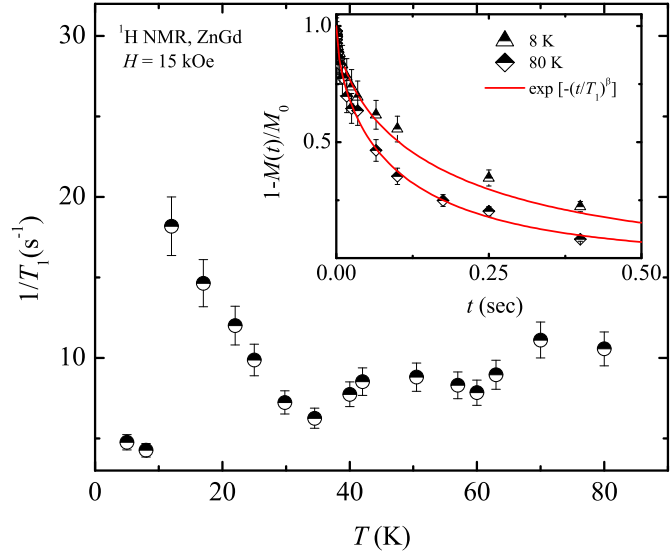

Fig.12. (Color online) Temperature dependence of spin-lattice relaxation rate ($T_1^{-1}$) in [ZnGd] in an applied magnetic field $H$ = 15 kOe. The inset shows the recovery of longitudinal magnetization in [ZnGd] at various temperatures, and the solid lines are fits to a stretched exponential function, i.e., $\exp[-(t/T_1)^\beta]$.

At temperatures below 20 K, we observe an anomalous increase of $1/T_1$ (see Fig. 12). This behavior could be associated with a slowing down of the correlation time due to inter-cluster coupling of the large $Gd^{3+}$ magnetic moments. It is noted that the two values of $1/T_1$ measured below 10 K in Fig.12 are obtained in a temperature region where the proton NMR line width becomes quite large and therefore are not reliable. The other indications of inter-cluster coupling are the decrease of $\chi_m T$ in Fig. 5 and the deviation of the proton line width from linearity with $\chi$ observed in [ZnGd] (not shown here). However, it is not clear how to reconcile this scenario with the lack of effects on the magnetization isotherm at 2 K (Fig. 6), which instead can be interpreted in terms of a simple paramagnetic behavior.

For [NiGd], we were unable to collect a good proton NMR signal at temperature lower than 300K, possibly because of the presence of the large moment that evolved due to the ferromagnetically coupled Ni and Gd ions, which leads to very short relaxation times.

# VI. SUMMARY AND CONCLUSIONS

To summarize, we have presented systematic magnetization and $^1$H NMR measurements on recently synthesized, rare earth based, magnetic clusters [NiNd], [ZnGd] and [NiGd].

The bulk magnetic susceptibility data in [NiNd] are indicative of an intra-molecular antiferromagnetic interaction between $Ni^{2+}$ and $Nd^{3+}$ ions via oxygen which leads to a dimer state at low temperature, whose exact nature we could not establish at present. On the other hand, the magnetization data in [NiGd] demonstrate that the spin ground state is a dimer with total spin $S=9/2$ due to the ferromagnetic interaction between $Ni^{2+}$ ($S_{Ni} = 1$) and $Gd^{3+}$ ($S_{Gd} = 7/2$).

The bulk magnetic susceptibility data in [ZnGd] are indicative of paramagnetic behavior with a large magnetic moment arising from spin-only $Gd^{3+}$. The system behaves like a paramagnet in almost all the temperature range with anomalies at low temperatures.

NMR provides a unique perspective of the static and dynamic magnetic properties at the local level. The $^1$H NMR line width scales with the bulk magnetic susceptibility in [NiNd] and [ZnGd]. The hyperfine coupling constant obtained is of the order expected for a dipolar coupling between protons and the magnetic moment of the $3d$ and $4f$ ions in the molecule. From $^1$H NMR measurements, it is observed that, in the high temperature regime, the dimer [NiNd] behaves like a paramagnet whereby the fluctuation rate is high, resulting in a low value of spin-spin and spin-lattice relaxation rates. However, at low temperature, an intra-dimer antiferromagnetic correlation between spins builds up, and the spin-fluctuations become slow resulting in an increase in $T_1^{-1}$ and $T_2^{-1}$ with a peak in $T_1^{-1}$. The dynamical magnetic susceptibility as obtained from NMR spin-lattice relaxation rates in [NiNd] suggests the existence of spin correlation induced by fluctuations of effective $3d$ and $4f$ spins in [NiNd] with a single correlation frequency $\omega_c \propto T^{3.5}$. This might be a signature of the life time broadening of the exchange coupled spins in [NiNd] mediated by spin-phonon interactions. The onset of magnetic correlation developed in [NiNd] in the intermediate temperature regime is so far the first such NMR investigation in $3d$-$4f$ complexes.

Surprisingly, even in the simple paramagnetic system [ZnGd], there is an increase (however weak) in $T_1^{-1}$ with decreasing temperature and a dramatic increase in the proton line width below 15 K. This might signal the onset of inter-cluster Gd interactions, which appear to be of antiferromagnetic type as also signaled by the decrease of $\chi_m T$ in the same temperature range.

In conclusion, the single molecule magnet (SMM) behavior is not limited to transition metal complexes, but also it is possible in 3$d$-4$f$ binuclear and polynuclear complexes. So far there are not many polynuclear 3$d$-4$f$ complexes available in view of their synthetic difficulties and other related issues. Our investigation addresses the SMM behavior in this class of molecular clusters, which may attract the design and investigation of such variety of polynuclear complexes with promising physical properties. Further microscopic and high field magnetization studies on single crystals of these complexes are highly desirable to determine the anisotropy energy and other interesting features of these molecules. A suitable theoretical framework explaining the spin-orbit coupling and exchange interactions at the origin of the energy spectra of the Stark levels still needs to be developed.

## Acknowledgements

We acknowledge support from the EU Network of Excellence MAGMANet. The work in Pavia was supported in part by a grant from the Italian Ministry of Education PRIN 2008. Work at the Ames Laboratory was supported by the Department of Energy, Basic Energy Sciences, under Contract No. DE-AC02-07CH11358. T. D. P. and M. A. acknowledge the CNCS for financial support (Grant No: 1912/2009).

---

* pkhuntia@gmail.com